\documentclass[useAMS, a4paper,fleqn,usenatbib]{mnras}

\usepackage{graphicx}	
\usepackage{amsmath}	
\usepackage{amssymb}	
\usepackage{amsfonts, txfonts}

\newcommand\mearth{{\,{\rm M}_{\oplus}}}
\newcommand\mj{{\,{\rm M}_{\rm J}}}

\newcommand\simgt{\ga}

\def\del#1{{}}

\title[]{Planets, debris and their host metallicity correlations}

\author[M. Fletcher, S. Nayakshin]{
Mark Fletcher\thanks{E-mail: mf240@le.ac.uk} and Sergei Nayakshin
\\
Department of Physics and Astronomy, University of
  Leicester, Leicester LE1 7RH, UK.
}

\date{Accepted 17 May 2016. Received YYY; in original form ZZZ}

\pubyear{2015}

\begin{document}
\label{firstpage}
\pagerange{\pageref{firstpage}--\pageref{lastpage}}
\maketitle

\begin{abstract}
Recent observations of debris discs, believed to be made up of remnant planetesimals, brought a number of surprises. Debris disc presence does not correlate with the host star's metallicity, and may anti-correlate with the presence of gas giant planets. These observations contradict both assumptions and predictions of the highly successful Core Accretion model of planet formation. Here we explore predictions of the alternative Tidal Downsizing (TD) scenario of planet formation. In TD, small planets and planetesimal debris is made only when gas fragments, predecessors of giant planets, are tidally disrupted. We show that these disruptions are rare in discs around high metallicity stars but release more debris per disruption than their low [M/H] analogs. This predicts no simple relation between debris disc presence and host star's [M/H], as observed. A detected gas giant planet implies in TD that its predecessor fragment was not disputed, potentially explaining why DDs are less likely to be found around stars with gas giants. Less massive planets should correlate with DD presence,  and sub-Saturn planets ($M_{\rm p} \sim 50 \mearth$) should correlate with DD presence stronger than sub-Neptunes ($M_{\rm p} \lesssim 15 \mearth$). These predicted planet-DD correlations will be diluted and weakened in observations by planetary systems' long term evolution and multi-fragment effects neglected here. Finally, although presently difficult to observe, DDs around M dwarf stars should be more prevalent than around Solar type stars.
\end{abstract}


\section{Introduction}\label{sec:intro}

\subsection{Planet debris and the classical scenario}\label{sec:classico}

Asteroids, comets and minor bodies in the Solar System are remnants from the
planet formation era \citep[see][for a recent review]{JohansenEtal14a}.  Solid
debris is also detected around a good fraction of nearby Solar type stars
\citep{Wyatt08} through thermal emission by grains in infra-red
\citep{OudmaijerEtal92,ManningsBarlow98}. These grains should have been blown
away rapidly by the stellar radiation pressure. Their continuos presence therefore
requires a large reservoir of much bigger solid bodies feeding the
fragmentation cascade \citep[e.g.,][]{Hellyer70} and producing the observed grains continuously. 
Naturally, observations of debris discs (DDs) are expected to shed light on planet formation theories.

Core Accretion theory \citep{Safronov72} stipulates that
planet formation starts with formation of minor solid bodies called
planetesimals.  By sticking together, planetesimals form terrestrial planet
mass "embryos", which grow further by continuing to accrete planetesimals
\citep[e.g.,][]{HayashiEtal85,Wetherill90,WetherillS93,KL99}. As solid cores reach the critical
mass of $\sim 10 M_\oplus$, they start to accrete gas \citep{Mizuno80,Stevenson82,Rafikov06}, which culminates in the
emergence of gas giant planets \citep[e.g.,][]{PollackEtal96,IdaLin04a,HubickyjEtal05}. Protoplanetary
discs around more metal rich stellar hosts are naturally expected to form more
massive planetesimal discs \citep{IdaLin04b}. This results in a more rapid
assembly of massive solid cores, and hence yields more gas giant planets in
higher metallicity environments
\citep[e.g.,][]{IdaLin04b,MordasiniEtal09b}.  The expected correlation chain in CA is thus
"higher [M/H] $\rightarrow$ more planetesimals $\rightarrow$ more
solid cores $\rightarrow$ more gas giants" .

\subsection{Metallicity correlation challenges}\label{sec:Z}

\subsubsection{Planets}\label{sec:Z_planets}

Historically, this correlation chain has been tested by observations in roughly the reverse order since gas giants are the "easiest" to observe. It is now well established that frequency of appearance of gas giant planets correlates strongly
with the host star's metallicity \citep[e.g.,][]{Gonzalez99,FischerValenti05,SanterneEtal15}. 
However, sub-Neptune planets are abundant at all metallicities
\citep[e.g.,][]{SousaEtal08,MayorEtal11,BuchhaveEtal12}. More recent analysis
of \cite{WangFischer15} reveals that planets with radii smaller than $4
R_\oplus$ (e.g., smaller than Neptune) are correlated with host's metallicity
albeit weakly: hosts with [M/H]$ > 0$ are twice more likely to have a planet
in this radius range compared with sub-Solar metallicity hosts (for gas
giants the corresponding number is $\sim
9$). 

  The weak correlation of massive cores with [M/H]  has been interpreted to imply that
  massive cores in metal-rich systems reach the critical mass for runaway gas
  accretion before the gas disc dissipates, so they quickly become gas
  giants \citep{IdaLin04b,MordasiniEtal09b}. In metal poor systems, in this picture, massive cores form in gas-free environment and hence do not make gas giants. 
  
  However, the observed solid cores outnumber gas giants by approximately ten to one
  \citep[e.g.,][]{MayorEtal11,HowardEtal12}.
  If cores do become gas giants very frequently in metal-rich systems, why
  are the gas giants not as numerous as massive cores? Furthermore, recent models of pebble accretion 
  suggest that cores may grow orders of magnitude faster than they do in the planetesimal-based paradigm
  \citep[e.g.,][]{OrmelKlahr10,LambrechtsJ12,LJ14}. Furthermore, if core accretion is to explain the rapid formation of massive cores at separations of $\simgt 10$~AU in the Solar System \citep[e.g.,][]{HB14}, and in the likes of HL Tau \citep[e.g.,][]{BroganEtal15,DipierroEtal15,PinteEtal16}, then we should accept that observations demand a rapid core formation even at large distances from the star. This casts doubts on the suggestions that cores grow after the gas disc is dissipated in metal-poor discs and leaves the observed planet -- [M/H] correlations somewhat puzzling in the context of CA.
 
 \subsubsection{Debris discs}\label{sec:dd_chal}
  
Debris disc  detection frequency does not correlate with [M/H] of host stars \citep{MaldonadoEtal12,MarshallEtal14,Moro-MartinEtal15}, directly contradicting a key assumption usually made about formation of planetesimals \citep[that they form in a greater abundance in higher metallicity hosts, e.g.,][]{IdaLin04b}. Furthermore, observed DDs also do not correlate with the
presence of gas giant planets \citep[e.g.,][]{MMEtal07,BrydenEtal09,KospalEtal09}. In fact, stars orbited by a gas giant are twice less likely to host a detected DD than stars orbited by planets less massive than $30 M_\oplus$ \citep{Moro-MartinEtal15}.

These observational results are somewhat paradoxical. All the planets referred to in these surveys are relatively close-in ones, e.g., orbiting the star at separation $a\lesssim$ a few AU. Spatial scales of the observed DDs are however much larger, e.g., tens of AU \citep{Wyatt08,Moro-MartinEtal15}. Thus the {\em observed} planets and DDs should not interact directly. As both populations should be more abundant at higher [M/H], it would imply that there should be a positive cross-correlation between planets and debris, contrary to what is observed.

\cite{RaymondEtal11} suggested an explanation for these observational results. They showed that mutual interactions of giant planets initially orbiting the host star at a few to $\sim 10$ AU distance are capable of exciting a very large velocity dispersion in the population of planetesimals located further out. This could fuel a strong fragmentation cascade, depleting the debris rings by $\sim 1$ Gyr age on which DDs are typically observed. These calculations also predicted a strong correlation between DDs and terrestrial-mass planets. \cite{WyattEtal12} report
a correlation between debris discs and presence of planets less massive than
Saturn, although the more recent study of \cite{Moro-MartinEtal15} casts
doubts on that correlation. 

The scenario suggested by \cite{RaymondEtal11} is physically reasonable, but the observed gas giants are very rare. The authors invoke three gas giants in a disc. Observations however show that there is less than $0.05$ giant planets per star within period of 400 days on average \citep{SanterneEtal15}. Microlensing surveys \citep{ShvartzvaldEtal15} constrain the number of gas giants beyond the snow line region to a similarly small regiment. Directly imaged surveys also find that massive gas giants in separations of $10-100$~AU orbit at most  a few \% of stars \citep[e.g.,][]{BillerEtal13,BowlerEtal15}\footnote{We note in passing that the rarity of {\em observed} gas giants in TD is much less of a problem because the huge majority of Jupiter mass gas fragments born in the outer disc are eventually disrupted and leave behind the much more abundant sub-giant planets. See \cite{Nayakshin16a} for more detail.}.
 In addition to this, the scenario of \cite{RaymondEtal11} predicts a strong decay in the dust luminosity as a function of stellar age as young massive DDs self-destruct in fragmentation cascades. Fig. 4 of \cite{Moro-MartinEtal15} does not show any obvious trend in DD detection frequency with time, adding doubt to this theoretical picture. 
 


\subsection{Tidal Downsizing alternative}\label{sec:TDalt}

Here we propose a solution to all of these observational paradoxes based on a different planet formation theory.
In Tidal Downsizing (TD) scenario
\citep{BoleyEtal10,Nayakshin10c}, planet formation begins with gravitational instability \citep[GI;
  e.g.,][]{Kuiper51,CameronEtal82,Boss97} of a massive young protoplanetary
disc when the latter hatches $\sim 1$ Jupiter mass gas fragments at distances
of tens to hundreds of AU \citep{Rice05,DurisenEtal07} from the host
star. These fragments migrate inward very rapidly
\citep[e.g.,][]{VB06,BaruteauEtal11,TsukamotoEtal14}. As they contract slower
than they migrate, most are disrupted by tides from the parent star. Grain
sedimentation within the fragments is believed to form massive solid cores in the
centre of the fragments
\citep{Kuiper51,McCreaWilliams65,HelledEtal08,BoleyEtal10,Nayakshin10b}. When
the fragments are disrupted, the cores are released back into the
protoplanetary disc, potentially yielding solid core planets
\citep{BoleyEtal10,ChaNayakshin11a}. Cores and gas fragments that managed to
avoid tidal disruption continue to migrate in and may end up arbitrarily close to the parent star \citep[see][]{NayakshinFletcher15}.

\del{\cite{NayakshinFletcher15} perform population synthesis of this scenario and find it promising in
explaining many of the observed properties of exoplanets at all separations.}

\subsubsection{Planet -- metallicity correlations}\label{sec:TD_planets}

\cite{Nayakshin15a} found that gas fragments contract rapidly when pebbles (grains of a few mm in size) are deposited in their outer envelopes by pebble accretion \citep{OrmelKlahr10,LambrechtsJ12} from the protoplanetary disc. When the fragments contract sufficiently, e.g., when their central temperature exceeds $\sim 2,000$ K, Hydrogen molecules in the fragment dissociate and it collapses to much higher density \citep{Bodenheimer74}. For a rapidly migrating fragment, gravitational collapse is the only way of avoiding an imminent tidal disruption in the inner few AU. Assuming pebbles are more abundant at high metallicities (note that this assumption has nothing to do with abundance of planetesimals, see below), one obtains higher pebble accretion rates, and hence a faster fragment collapse at high [M/H]. Unsurprisingly then, one obtains a strong positive metallicity correlation for gas giant planets in the inner few AU \citep{Nayakshin15b,Nayakshin15c} because only the planets that managed to collapse faster than they migrated into that region survive.

On the other hand, \cite{NayakshinFletcher15} show that sub-giant planets do NOT follow a strong [M/H] correlation because these planets are formed when the gas fragments are disrupted. There is however no simple anti-correlation between the low mass planets and the host's metallicity. More massive cores are made inside more metal rich fragments. Therefore, while solid cores are most abundant {\em by numbers} at low [M/H], the most massive of them are found at high [M/H]. The combination of these two effects does not produce a clear cut correlation with metallicity of the host star, qualitatively as observed.


\subsubsection{Origin of debris discs in TD}

Astrophysical existence of Asteroid and Kuiper belts in the Solar System and DDs around other stars is usually taken as a proof that planetesimal synthesis has taken place as predicted by the planetesimal hypothesis \citep{Safronov72}.

 However,  \cite{NayakshinCha12} showed that TD can also naturally yield DD-like structures. In particular, it was suggested that, as the central regions of the fragments become dominated by grains rather than by H/He gas, gravitational collapse of the solid component may follow \citep{Nayakshin10a}, as in the model of \cite{GoldreichWard73}, except for the different geometry.
Simulations show that self-gravitating gas fragments formed in proto-planetary discs always rotate rapidly \citep[e.g.,][]{MayerEtal04,BoleyEtal10,GalvagniEtal12}, so that not all solids are likely to condense into  a single central core due to excess angular momentum. Fragments larger than $\sim 1-10$~km decouple from the gas
aerodynamically, that is, the timescale for in-spiral of these bodies
into the core is $\simgt 10^5$~years, which is longer than the expected
lifetime of the host fragments \citep[see Fig. 1 in][]{NayakshinCha12}. 

The issue was further studied in Nayakshin (2016b). When the grains collect in the central region of the fragment and start to dominate the volume density there, they can collapse due to their own self-gravity, separating out from the H/He gas \citep[somewhat like in the model of][except for a different geometry]{GoldreichWard73}. It turns out that regions of all linear sizes collapse on the same time scale. This means that if there are perturbations seeded by convection, turbulence or clump rotation, they can grow and collapse into separate solid bodies. However, only bodies large enough are able to decouple dynamically from the overall collapsing flow. This sets a minimum body size of about 1 km.

Hydrodynamical simulation of a fragment disruption showed  that large solid bodies closest to the core remain bound
to it, perhaps contributing to formation of satellites (as needed for Neptune
and Uranus). Bodies farther out are however unbound from the core when the gas
is removed. After a few rotations around the star, the disrupted solids are arranged into debris rings with kinematic properties (e.g., eccentricity
and inclination) resembling the Kuiper and the Asteroid belts in the Solar System.

One can immediately see that TD naturally explains why planets observed at small separations may be connected to DDs observed at tens to hundreds of AU. The close-in planets migrated through these far away regions before arriving at their present day locations. Disruption of a planet at large distances (a) creates a DD there, and (b) changes the type of planet (by downsizing it into a lower mass planet) that may be observable at close separations to the star, provided that the planet migrates there before the gas disc is dissipated. It is also intuitively clear that the DD-planet-metallicity relations predicted by the TD scenario must differ from those of the CA theory, and it is hence important to clarify these to aid future observational testing of these two planet formation scenarios.

This paper is structured as following. In \S 2 we present the numerical methods with which we shall test how predictions of TD scenario for formation of both planets and debris discs depend on  the host star's metallicity. In \S \ref{sec:planets} we explore the results for planets, and in \S \ref{sec:debris} the focus is on the debris discs. Section 5 presents a brief discussion and a comparison of the theory and observations. We conclude in \S 6.

%
%
\section{Numerical Method}\label{sec:method}

Our aim is to understand how metallicity of the protoplanetary disc, assumed below to be equal to that of the host star, influences the fate of a gas fragment formed in the outer disc by gravitational instability. Since fragments may be hatched at a range of separations in discs with varied initial conditions, it is best to approach this question statistically, by performing a population synthesis study \citep[cf. similar approach for CA theory, e.g.][]{IdaLin04a}. To this end we use the same code and general setup as used by \cite{NayakshinFletcher15} and \cite{Nayakshin16a}, briefly overviewing it here. 

Note that our main results, as stated in the Abstract, are very insensitive to the parameter choices in the population synthesis. These results are driven by the tidal disruption of pre-collapse fragments, the central pillar of the Tidal Downsizing scenario \citep{Nayakshin10c}, and depend on parameter choices only quantitatively but not qualitatively. This is true as long as pebble accretion onto precollapse fragments is sufficiently rapid to dominate their effective cooling \citep{Nayakshin15a,Nayakshin15b}, or else Tidal Downsizing scenario fails to reproduce many of the observations \citep[e.g.,][]{ForganRice13b}.

\subsection{Population synthesis}\label{sec:ps}

 A 1D viscous time dependent code is used to evolve in time the disc
surface density, $\Sigma$, midplane temperature and other disc properties. A logarithmic grid
extending from the inner radius, $R_{\rm in}=0.08$~au, to the outer radius,
$R_{\rm out}=400$~au, is used.  The disc initial surface density is proportional to
$1/R$ with an exponential rollover at $R_0 = 100$~AU. The total disc mass is
randomly chosen between the limits shown in Table 1 (note that when $M_{\rm d} \approx 0.15 M_\odot$ the Toomre parameter $Q$ drops to $\sim 1.5$ at $R\sim 80-90$~AU, so that the disc can fragment there). 

The disc viscosity parameter, $\alpha$, is fixed for
each simulation but its value is drawn randomly in the logarithmically uniform
fashion between the minimum and the maximum values shown in Table 1. The disc is
photo-evaporated with the overall normalisation of the photo-evaporation rate
also being a Monte-Carlo variable ($\zeta_{\rm ev}$), whose boundaries are
selected to fit the roughly exponential decrease in the disc fraction with
time \citep[see figure B1 in][]{Nayakshin16a}. In addition to the viscous evolution, our disc is also affected
by the gravitational torques that drive planet migration and may open gaps in the disc \citep[e.g.,][]{GoldreichTremaine80,LinPap86}. The gas disc physics and
assumptions are therefore quite similar to those employed by CA population
synthesis authors \citep[e.g.,][]{MordasiniEtal09b}. Most of the distinctions with the CA modelling is in
the planet formation physics rather than in the gas disc physics.

A population synthesis model is started with a fragment of an initial mass, $M_0$, in the
range between $1/3 \mj$ and $8\mj$, placed in the disc at distance $a_{\rm pl}$
between $70$ and 105~au (where the disc Toomre's parameter has a broad
minimum). 

 \cite{ForganRice13b} calculate the mass of the gas fragments, $M_0$, based on analytical estimates for Jean's mass ($M_{\rm jeans}$) in the their self-gravitating disc. While this approach is in principle superior to ours, which decouples $M_0$ from the conditions in the disc, we feel that any analytical calculation of $M_{\rm jeans}$ cannot be deemed reliable at this point. The fundamental problem here is not how $M_{\rm jeans}$ is derived, but the approximations on which the standard accretion disc theory is built upon. In particular, \cite{Shakura73} approach utilises vertically integrated, that is, vertically averaged, equations for the disc. This approach cannot be more accurate than a factor of $\sim 2$, as pointed out by \cite{Svensson94}, who retained the uncertainty in accretion disc equations (see their parameter $\zeta$). Furthermore, the location of the disc fragmentation and disc properties strongly depend on the irradiation of the disc by the background and stellar fluxes, and these can vary widely at same disc mass \citep[e.g.,][]{StamatellosEtal11}. In addition, simulations show that presence of one fragment in the disc may cause the disc to fragment in locations/conditions where it would normally not fragment \citep{Meru15}, clearly showing that analytical methods cannot capture the whole variety of outcomes for fragmenting discs.

Since the mass of the fragments depends on the disc scale-height, $H$, as $M_{\rm jeans}\propto H^3$ \citep[e.g., eq. 6 in][]{KratterEtal10}, an uncertainty in $H$ by a factor of $\sim 2$ leads to an order of magnitude uncertainty in $M_{\rm jeans}$. As debates about the exact nature of fragmentation in gravitationally unstable protoplanetary discs continue to rage \citep[e.g.,][]{YoungClarke16,KratterL16}, we feel that it is safer to assume that the initial fragment mass varies in a broad range, and explore the resulting outcomes. We do not consider fragments more massive than $8\mj$ because these are not likely to be tidally disrupted as they contract rapidly and so they do not produce debris discs. The main conclusions of this paper are qualitatively independent from the initial fragment mass distribution as long as fragments with mass $M_{\rm p} \sim 1\mj$ are present in the distribution\footnote{We also note that initial conditions of the classical Core Accretion theory are even more uncertain than ours, as the planetesimal size considered varies from 1 to over 100 km. While such uncertainties are unfortunate one can still learn a great deal from the calculations and hope that initial conditions will be better constrained in the future \citep[e.g.,][]{Wetherill90}.}.

Once the fragment is born in the disc, its separation starts to decrease due to the tidal torques from
the disc, and the calculation proceeds until the disc mass reduces to zero due
to accretion onto the star and photo-evaporation. The end result of a
calculation comprises the final location, mass and composition of the planet.

We use a 1D spherically symmetric approximation
for the fragment. Gas fragments are allowed to
accrete grains a few mm in size from the disc via pebble accretion in the
Hills regime \citep{LambrechtsJ12} unless the planet opens a gap. Gas
accretion onto the planet is assumed inefficient
\citep[see][]{NayakshinCha13}. Pebbles are deposited in the outer layers of
the planet. Both incoming pebbles and the grains present in the fragment at
its birth grow via sticking collisions as they sediment
\citep[e.g.,][]{Boss97}. Three grain species (water, rocks and CHON) are considered. See
sections 4.2, 4.3, and 4.6 of \cite{Nayakshin15c} for further detail of grain
treatment.

The grains that settle all the way to the centre of the fragment are accreted
onto the core \citep[see][]{HS08,HelledEtal08,Nayakshin10b}. The internal
structure of the core is not modelled, assuming instead a sphere with a fixed
material density, $\rho_0 = 3$~g~cm$^{-3}$. As grains accrete onto the core,
their accretion luminosity is released back into the gaseous envelope of the
fragment, which produces important feedback effects for core masses exceeding
a few $\mearth$ \citep{Nayakshin16a}.

Table \ref{tab:IC} summarises the initial conditions used for the population
synthesis models. As noted they are almost identical to those from
\cite{Nayakshin16a}. In the table, $f_{\rm m}$ is the type I migration
multiplier, which multiplies the type I migration time scale, and $v_{\rm br}$ is the breaking velocity of grains within the
fragment (grain collisions at velocities higher than that lead to grain
fragmentation rather than growth).  The mass fraction of pebbles, e.g., grains of moderately large size \citep{OrmelKlahr10,JohansenLacerda10}, here set to $\sim 3$ mm, with respect to the total grain content of the protoplanetary disc, is $f_{\rm peb}$. The metallicity [M/H] distribution of our
host stars is a gaussian with a mean of 0 and a standard deviation of 0.22. The exact values of the population synthesis parameters do not influence our main results significantly. 30,000 population synthesis runs are performed in this study.

\begin{table}
\caption{Parameters of the population synthesis model and their values. See text for detail.}
\centering
\begin{tabular}{c c}
\hline \hline
Parameter & Range \\
\hline
$M_{\rm disc}$ [$M_{\odot}$]  &  0.075 - 0.2 \\
$M_{\rm p}$ [$\mj$] &  0.333 - 8 \\
$a_{0}$ [AU]  &  70 - 105 \\
$\alpha$  &  0.005 - 0.05 \\
$f_{\rm m}$  &  1 - 4 \\
$v_{\rm br}$ [m s$^{-1}$]  &  5 - 30 \\
$f_{\rm peb}$ & $0.02 - 0.04$ \\
\hline
\end{tabular}
\label{tab:IC}
\end{table}

\subsection{Debris model}\label{sec:debris_mass}

It is not possible to simulate particle instabilities that may form large solids in the central part of the fragment within 
our 1D spherically symmetric model for the planet. We
therefore explore two different assumptions about the mass of the debris
formed in the fragment which we believe roughly bracket the possible outcomes. We shall later see that these two assumptions lead to same conclusions as far as debris disc -- planets -- metallicity correlations are concerned.

First of all, one can argue that the total mass of solid debris made by the
fragment is a small fraction $0< \zeta \ll 1$ of the total mass of all of the solids
inside the planet (note that we use "planet" and "fragment" inter-exchangeably in this paper), that is,
\begin{equation}
M_{\rm deb} = \zeta Z M_{\rm p}\;,
\label{zeta}
\end{equation}
where $Z$ is the metallicity of the planet at the moment of its tidal
disruption. This assumption is reasonable but ignores the fact that grains
sediment differently in fragments of different temperature or age. For
example, none of the three grains species can reach the central part of a fragment with central temperature 
$T_{\rm c}
\simgt 1500$~K, since they are vaporised. The grains  hence cannot be separated from the gas to make large
solids, whereas in a fragment with central temperature of, e.g., $500$ K, refractory grains will sediment but organics and water ice will not \citep[e.g.,][]{HelledEtal08}. Furthermore, even if grains are not vaporised, they sediment at a finite velocity, and this process is opposed by the convective mixing of grains \citep[e.g.,][]{HB11}, so the fragment's age at the point of its tidal disruption is important in determining the fraction of grains that sedimented into the centre.

The second assumption one can make is that the total debris mass released back into the disc is a
fraction of the core mass,
\begin{equation}
M_{\rm deb} = \beta M_{\rm core}\;,
\label{beta}
\end{equation}
where $\beta$ is a positive number. This model is more reasonable from the
point of view of grain physics since core growth in our model does include
grain vaporisation at high temperatures and other grain sedimentation
microphysics. However we are still unable to constrain $\beta$ since we do not
know the angular momentum of our fragments, how it is distributed within them,
and where exactly large solids form.

While these two models leave normalisation of the debris disc {\em mass} dependent
on either $\zeta$ or $\beta$, the metallicity trends of our models should be
independent of the exact values of these free parameters as long as $\zeta$ or
$\beta$ are independent of the host star's metallicity. It is difficult to
see, for example, why the distribution of angular momentum within the fragment
would be a strong function of metallicity of the disc, and hence why $\beta$
would be a function of [M/H]. 


To make our model qualitatively consistent with debris disc observations in terms of the typical DD mass, we set  $\beta=1/10$, and $\zeta=1/250$. With these choices, our DD masses are typically $M_{\rm deb} \lesssim 1 \mearth$ {\it per disrupted fragment}. Since we expect a dozen or so fragments per a typical DD, the overall mass is expected to be a few to $10 \mearth$. This is qualitatively consistent with fig. 3 of \cite{Wyatt08}, which shows the observed {\it dust} masses. In the protoplanetary disc phase, time $t\lesssim 10$ Million years, the DD mass is likely to be smaller than the mass of the dust in the disc, which are $\sim 10 - 100\mearth$ in the figure. Also, we have the minimum mass limit on $M_{\rm deb}$ from the older discs which had their gas-dust disc depleted, which requires $M_{\rm deb} \gtrsim 1 \mearth$ \citep[see fig. 3 in][at $t\gtrsim 10$ Million years]{Wyatt08}. Therefore, typical DD masses are probably in the range of a few to $\sim 10\mearth$.

Finally, we need to make some distinction between low and high mass debris discs since observationally the latter are of course more likely to be observed. We model this observational detection bias with a toy prescription in which the probability of detecting a debris
disc of mass $M_{\rm deb}$ is
\begin {equation} 
P\left(M_{\rm deb}\right) = \frac{M_{\rm deb}^{2}}{M_{\rm deb}^{2} + M_{\rm det}^{2}} \;,
\label{detection}
\end {equation} 
where $M_{\rm det}$ is a detection mass limit. For example, if $M_{\rm deb} = M_{\rm det}$, only half of the debris discs produced by the population synthesis are detectable in our mocked observation. Below we chose several values of $M_{\rm det}$ to investigate how our conclusions depend on sensitivity of our synthetic "observation" of population synthesis results since it is not immediately clear what the right value of $M_{\rm det}$ is.

%
%

\section{Results: Planets}\label{sec:planets}

\begin{figure*} 
\includegraphics[width=6in]{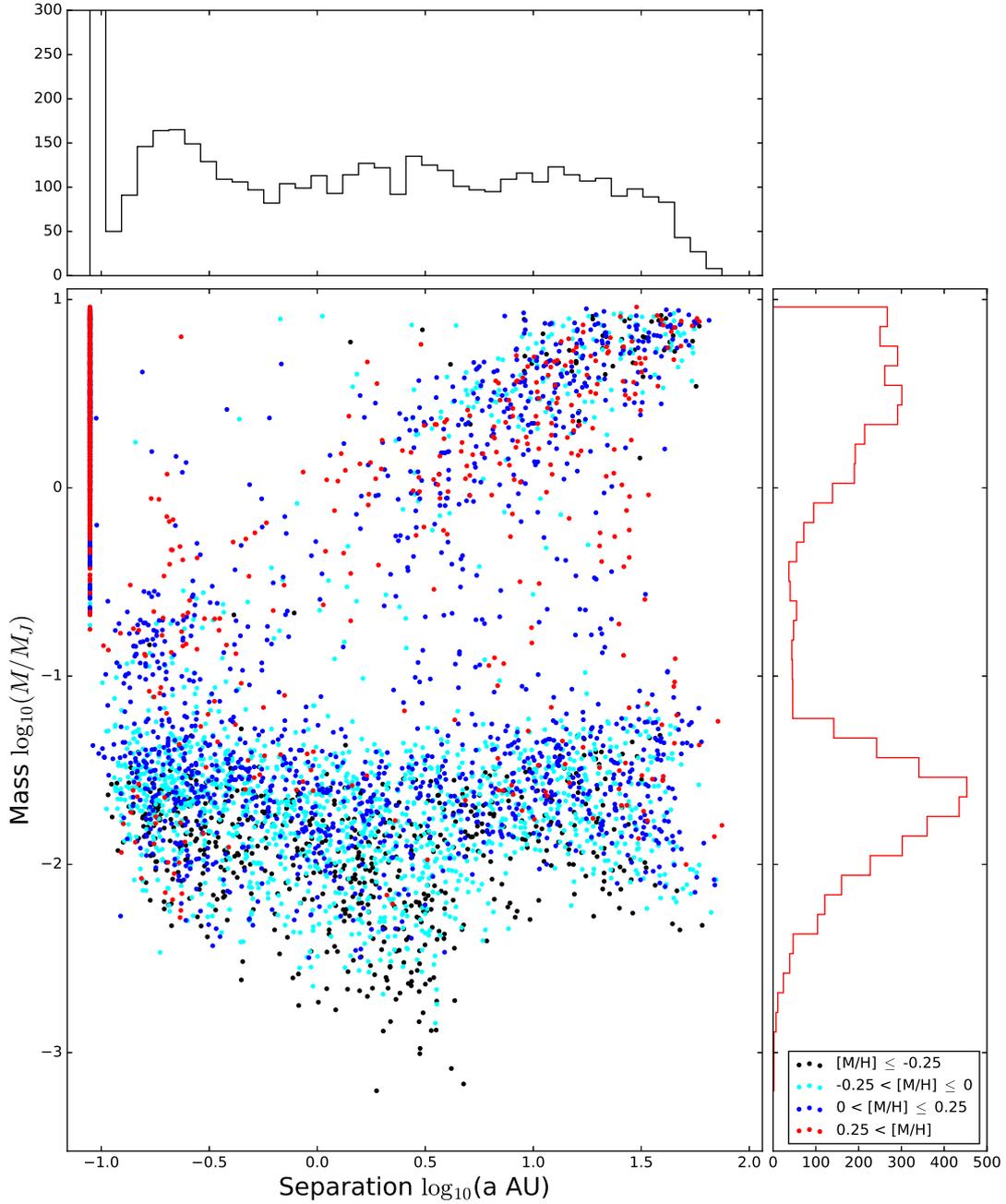}
\caption{Planet mass versus planet separation plane from the population synthesis
  calculations.  The colour of the points indicates metallicity of the host star, grouped in four bins as
  shown in the legend. The histograms above and to the right of the main
  figure show the distribution of the planets over the planet's separation and mass,
  respectively. Note a strong role of host star's metallicity in the inner few AU disc for $\sim 1$ 
  Jupiter mass planets but much less so for Super-Earth mass planets.}
  \label{fig:scatter1}
\end{figure*}

Fig. \ref{fig:scatter1} shows planet mass versus its separation
obtained in the population synthesis calculations. The colours of the symbols
indicate the metallicity of the parent star, grouped in four bins  as detailed in the
legend. Only $1/5$ of the actual 30,000 population synthesis runs are plotted
for clarity, with systems shown in the figure selected randomly and uniformly from the
total set of results. 
The histogram above the main panel in the figure shows the distribution of the planets over the planet-star separation, whereas the histogram on the right shows the planet mass function integrated over all separations.

These results are very similar to those presented earlier in \cite{NayakshinFletcher15} and \cite{Nayakshin16a}. In the inner few au, there is not a single gas giant planet with mass $M_{\rm p} \sim 1\mj$ in the lowest metallicity bin (black). The [M/H] distribution of these planets is clearly shifted towards positive [M/H], qualitatively as observed \citep[e.g.,][]{FischerValenti05}. Planet less massive than that, $M_{\rm p} \lesssim 0.1\mj$ or so, on the other hand, are distributed broadly over the metallicities, also in concord with observations \citep{MayorEtal11,HowardEtal12}.

It is instructive to look at probabilities, $p$, of different outcomes as a function of the stellar metallicity, [M/H]. To arrive at these, we count the initial number of fragments injected in the outer disc in the beginning of the simulations, $N_{\rm init}$, for each metallicity bin. We then count the number of planets that satisfy a given condition at the end of the simulations selected, $N_{\rm end}$,  e.g., the planet is a gas giant and is located in a specified separation range. The corresponding probability is then defined as
\begin{equation}
p = {N_{\rm end}\over N_{\rm init}}\;,
\label{p}
\end{equation}

\begin{figure} 
\includegraphics[width=1.0\columnwidth]{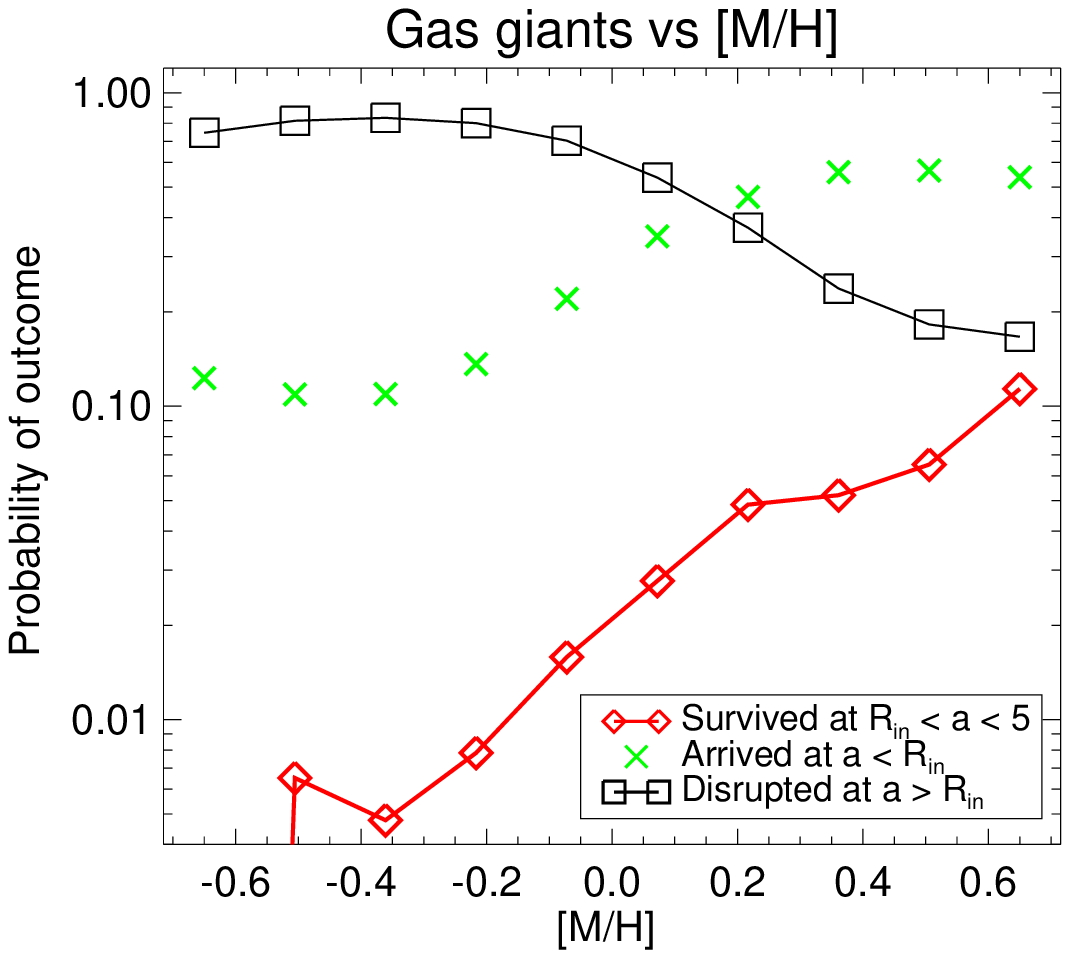}
\caption{Probability for a fragment to be tidally disrupted (black squares),
  reach the inner disc radius, $R_{\rm in}$ (green crosses), or survive in the
  inner region of the disc (red diamonds), all as a function of the parent
  star's metallicity [M/H]. Note that fragments are tidally disrupted most
  often at low [M/H], and least frequently at high [M/H]. This yields a
    positive gas giant planet -- host metallicity correlation of TD.}
\label{fig:giants_outcome}
\end{figure}

Three outcomes of the simulations are especially relevant to our paper, and the probabilities of these are shown in Fig. \ref{fig:giants_outcome}. The probability of a fragment to be tidally disrupted anywhere in the disc is shown with the black curve (squares). These disruptions leave behind dense planetary cores and also solid debris as described in \S \ref{sec:debris_mass}.

Next, the fragments that manage to collapse and avoid tidal disruption are split into two groups: those that continue to migrate and eventually arrive at the inner disc radius (the green crosses), $R_{\rm in}$, and those that are left "stranded" in the inner disc, $R_{\rm in} < a < 5$~AU, when the protoplanetary disc is depleted away (red diamonds). The young giant planets from the former group are likely to be disrupted at $R < R_{\rm in}$ since the disc is very hot there. Alternatively, these fragments may be driven all the way into the star and be assimilated there. Presumably only a small fraction of these planets survive as hot Jupiters, which are very rare observationally \citep{SanterneEtal15}. When disrupted, these fragments do not release any solid debris in our model since they are very hot, with their central temperature exceeding $\sim 30,000$~K soon after collapse \citep[see, e.g.,][]{Bodenheimer74}.

The planets that are left at $R_{\rm in} < a < 5$~AU when the disc dissipates away (the red
diamonds in the figure) are observable with the transit and radial velocity
observations. These show a strong positive correlation with [M/H] of the host
star \citep{Nayakshin15b}, as observed
\citep[e.g.,][]{FischerValenti05,MayorEtal11}. This result is due to pebble
accretion on pre-collapse gas fragments. Pebble accretion increases the weight of
such fragments and acts as an effective cooling mechanism that replaces the
inefficient radiative cooling, accelerating fragment contraction and eventual
collapse \citep[see][for the physics of this effect]{Nayakshin15a}.  At higher
[M/H], pebble abundance in the disc is higher, which increases pebble
accretion rate onto fragments. A larger fraction of gas fragments collapse
before they are tidally disrupted, and a greater fraction of gas giant planets
penetrates inside the "exclusion zone" at a few AU distance from the star
\citep[see the red line in fig. 3a in][]{Nayakshin16a}.

The corollary to the positive gas giant -- [M/H] correlation in our model is
that tidal disruption events at low [M/H] are more frequent than they are at high metallicities. This is apparent in
the behaviour of the black curve in fig. \ref{fig:giants_outcome}. This trend
shapes one of the key results of our paper, suggesting that, with other things
being equal, there should be more debris produced at low [M/H] discs than in
high [M/H] discs.

\begin{figure} 
\includegraphics[width=1.0\columnwidth]{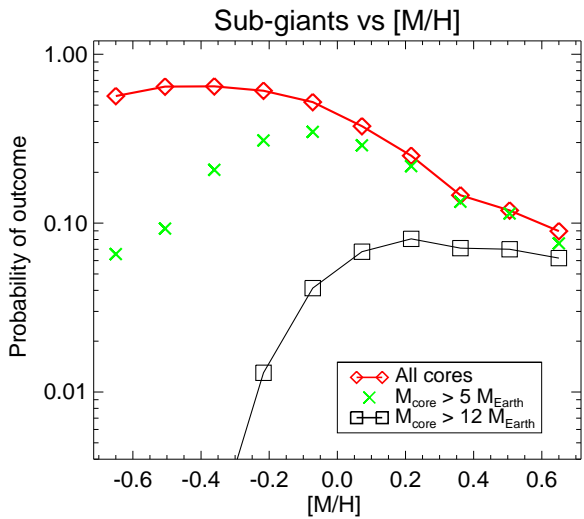}
\caption{The probability of core-dominated planet surviving in the inner 5 AU after the gas disc is dispersed in the
  population synthesis. The red curve shows all the cores while the green and
  the black are for cores more massive than 5 $M_{\oplus}$ and 12
  $M_{\oplus}$, respectively. Note that low mass cores are formed more readily
  at low [M/H] whereas high mass cores are formed more frequently at high
  [M/H] hosts.}
\label{fig:se_outcome}
\end{figure}

The probability of core-dominated planets being present in the inner 5 AU as a
function of metallicity is plotted in fig. \ref{fig:se_outcome}. The red diamonds show
this probability for all core masses. This curve is very similar to the black
curve from fig. \ref{fig:giants_outcome}. The green crosses and the black
square curves show probability of core formation for masses greater than
$5\mearth$ and $12 \mearth$, respectively. These clearly show that most
massive cores are made in the most metal rich environments, as could have been
expected, but in terms of numbers of all cores, there are more cores around low [M/H] hosts.

\del{We note in passing that this picture -- more cores by numbers at low
[M/H] but most massive cores appearing at higher [M/H] -- may be consistent
with the observed weak metallicity correlations for planets with radius
smaller than $4 R_\oplus$ \citep{WangFischer15}, although a quantitative comparison utilising
synthetic observations of our planets is yet to be made.}

%
%
%
%

\section{Results: debris discs}\label{sec:debris}

\subsection{Debris disc masses}\label{sec:dd_hist}

\begin{figure} 
\includegraphics[width=1.0\columnwidth]{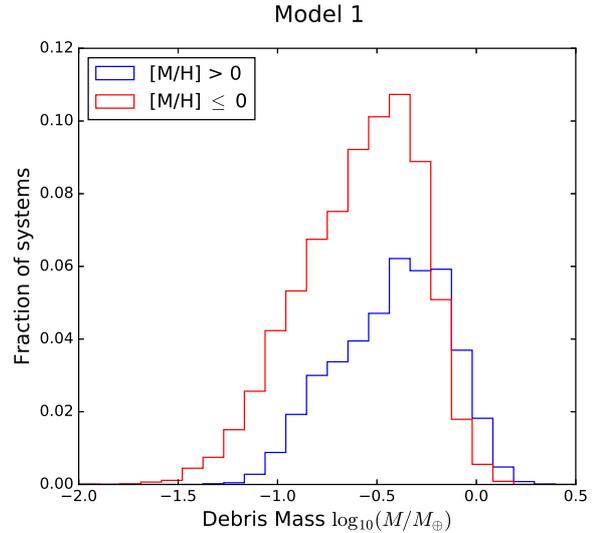}
\caption{The histogram of debris discs masses from our population synthesis calculations in Model 1, in which fraction $\zeta = 1/250$ of the fragment's metal content is released back into the disc as large solid debris bodies. As expected from fig. \ref{fig:se_outcome}, metal-poor stars (red histogram) host more DD systems but they are less massive on average compared with DDs around metal-rich stars (blue).}
\label{fig:frag_m_debris}
\end{figure}

Fig. \ref{fig:frag_m_debris} presents the distribution of the debris mass
obtained in the population synthesis via Model 1 (see \S
\ref{sec:debris_mass}), with $\zeta = 1/250$. We separated the histogram on the
metal rich, [M/H]~$>0$ (blue colour), and the metal-poor populations,
[M/H]~$<0$ (red). The peak of the histograms is at $M_{\rm c} \approx 0.35
\mearth$ for the former and $M_{\rm c} \sim (0.5-0.8) \mearth$ for the latter. As
expected, debris released by disruptions of metal-rich population is more
massive than that of [M/H]~$<0$ discs. As also expected based on
figs. \ref{fig:giants_outcome} and \ref{fig:se_outcome}, the low metallicity
environments produce more debris disruption evens per fragment.
The wide spread for the debris mass in fig. \ref{fig:frag_m_debris} is enhanced by the wide
range in the fragment's mass in the initial conditions for the simulations
(from $M_0 = 1/3 \mj$ to $M_0 = 8\mj$, see Table 1).

\begin{figure} 
\includegraphics[width=1.0\columnwidth]{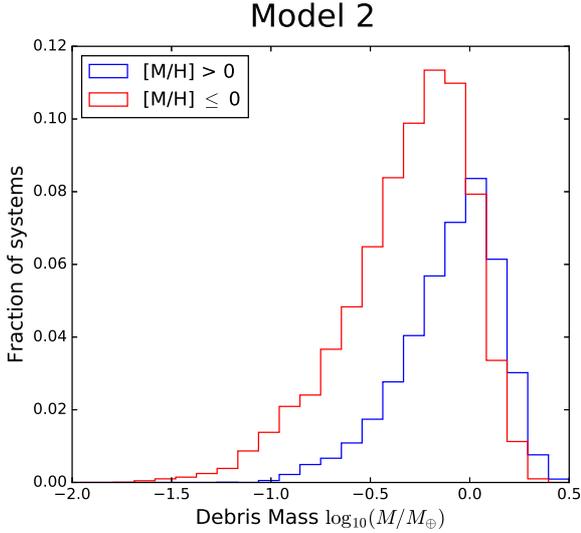}
\caption{Same as fig. \ref{fig:frag_m_debris} but for Model 2, in which the
  mass of the debris released is equal to $1/10$ of the core. Note that qualitatively the result is the same: metal rich systems produce $\sim $ two to three times more debris mass than their lower metallicity brethren. }
\label{fig:m_debris}
\end{figure}

As noted in \S \ref{sec:debris_mass}, Model 1 is an over-simplification which
does not take into account grain growth and sedimentation physics. In Model 2,
we instead assume that the mass of the debris released is $1/10$ of the mass of
the core, assembly of which does take into account grain
physics. Fig. \ref{fig:m_debris} shows the debris mass histograms for Model
2. This shows that, although there are important quantitative differences
between Model 1 and Model 2, qualitative trends with metallicity remain
similar. The mass of debris per disruption is still larger at higher
metallicity. However, the histogram for Model 2 is narrower than that for
Model 1, and one also notices a sharp roll-over at the high mass end. The
rollover is due to the feedback unleashed by massive cores on their fragments,
investigated in \cite{Nayakshin16a}. When the core mass exceeds $\sim 10\mearth$,
its luminosity output puffs up the pre-collapse fragment, slowing down grain
sedimentation and even disrupting the fragment in extreme cases. This leads to
a saturation of core masses at around $10\mearth$. Since our debris discs attain the mass of 10\% of $M_{\rm core}$ in Model 2, our DDs peak in masses around $1 \mearth$.

\subsection{Debris discs metallicity correlations}\label{sec:dd_vs_z}

\begin{figure} 
\includegraphics[width=1.0\columnwidth]{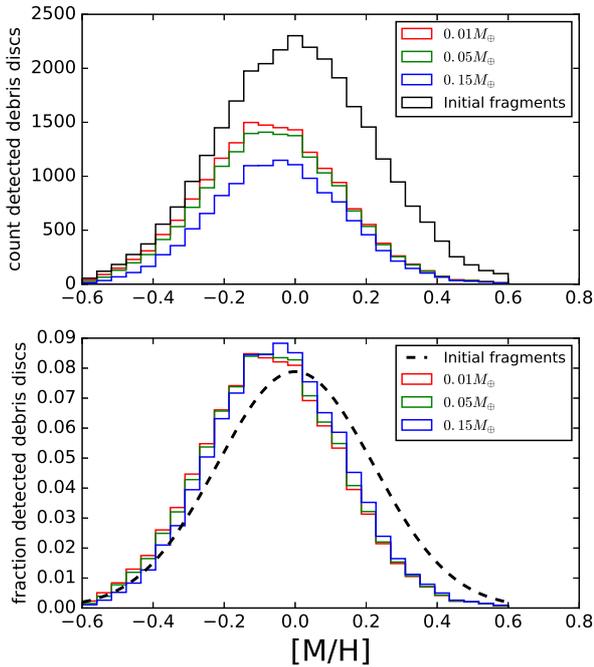}
\caption{The distribution of the detectable debris discs over metallicity of
  the host for Model 1 and three detection values of detection threshold,
  $M_{\rm det} = 0.1 M_{\oplus}$, $0.5 M_{\oplus}$ and $1.5 M_{\oplus}$, as
  indicated in the legend. The green histogram shows the initial fragment
  distribution over metallicities. The top panel shows the absolute number of
  systems whereas the bottom panel shows the histograms normalised to the
  integral of unity. }
\label{fig:hist_dd_model1}
\end{figure}

\begin{figure} 
\includegraphics[width=1.0\columnwidth]{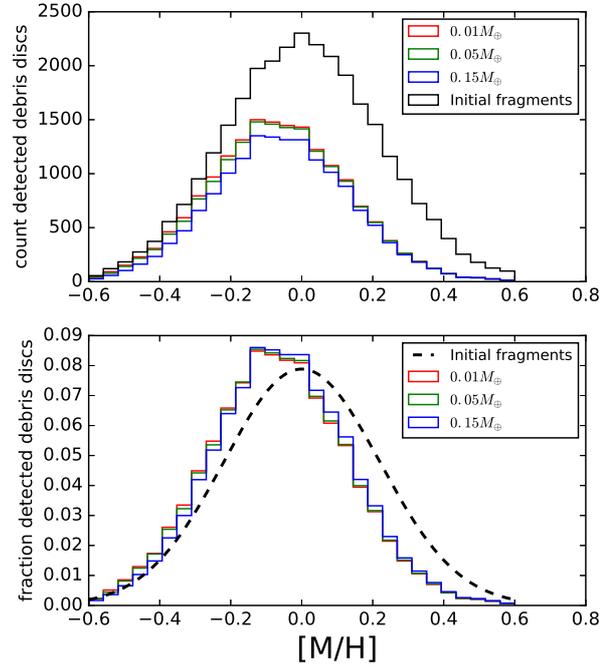}
\caption{Same as figure \ref{fig:hist_dd_model1} but for Model 2, in which the mass of the debris is given by given by $1/10$ of the core's mass.}
\label{fig:hist_dd_model2}
\end{figure}

Fig. \ref{fig:hist_dd_model1} shows the metallicity distribution of the
detectable debris discs for Model 1, where the debris mass is set to 
fraction $\zeta=1/250$ of the total metals' mass inside the disrupted
fragment. Three detection limits are considered as indicated in the legend in
the figure. The bottom panel shows the normalised distribution for the three
cases and also the gaussian distribution for the host star's metallicity
distribution.

The main conclusions from fig. \ref{fig:hist_dd_model1} are: (1) the
metallicity distribution for the detectable debris discs is similarly broad to
that of the host stars, with only mild skewness; (2) the shift in the [M/H]
distribution of debris disc hosts is to lower metallicities at low detection
threshold and toward higher metallicities for larger values of $M_{\rm
  det}$. These results are natural in the context of the discussion in \S
\ref{sec:planets}. At low [M/H], more gas fragments are disrupted, but they
have a relatively low metal content. Gas fragments at higher [M/H] are
disrupted rarely but they contain more metals. Therefore, if debris detection
threshold is low, the debris disc population will be dominated by the
metal-poor part of the population. In the opposite case, rare but metal-rich
disruptions at higher [M/H] are the main contributors.

Fig. \ref{fig:hist_dd_model2} shows a similar calculation but now assuming that the
mass of the debris is $1/10$ of the core (Model 2 in \S
\ref{sec:debris_mass}).  It is pleasing to see that main results are hardly
changed from those obtained with Model 1, suggesting that the insensitivity of
the debris mass to the host star metallicity that we find here is robust.
Since Model 2 is nevertheless more physically complete as discussed in \S
\ref{sec:debris_mass}, we shall continue our analysis based on just this
model.

\begin{figure} 
\includegraphics[width=1.0\columnwidth]{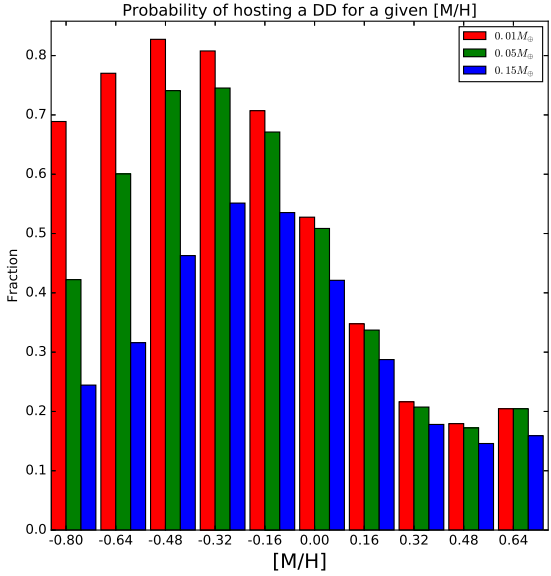}
\caption{Probability for a star of given metallicity to host a detectable debris disc for three values of  $M_{\rm det}$ (cf. eq. \ref{detection}), as specified in the legend. Note that sensitivity of our toy synthetic observations model (the value of $M_{\rm det}$) affects the prevalence of DDs as a function of metallicities.}
\label{fig:bar_all_z_frac}
\end{figure}

Fig. \ref{fig:bar_all_z_frac} summarises these results in terms of probability
that a host star of a given metallicity will have a detectable debris disc for
the three detection limits considered. The trends discussed above are clearly
visible but depend quite strongly on the sensitivity of our toy synthetic observation model.
A sensitive observation would have a low $M_{\rm det}$, so would detect debris discs mainly around low [M/H] stars, as seen from the red histogram. Brighter debris discs reside around increasingly more metal rich stars. The middle value for $M_{\rm det}$ shows in fig. \ref{fig:bar_all_z_frac} hardly any preference for [M/H] of the host. However, at the highest value for $M_{\rm det}$ considered, one finds debris discs only around high [M/H] stars. This is qualitatively consistent with the results of \cite{Moro-MartinEtal15} who found a lack of bright debris discs around low metallicity hosts.

We suggest that one should not read too much into the normalisation
of debris disc mass in our plots. Our synthetic observation model can be at best described as a
toy model, and the detection limits chosen are arbitrary. Furthermore, here we
considered just one fragment migrating per star, but in reality one may expect
a dozen to form per lifetime of the disc \citep[e.g., see simulations
  by][]{VB06,BoleyEtal10,ChaNayakshin11a}. However, we believe that the
metallicity trends obtained here will remain qualitatively same in a more
sophisticated analysis with many migrating fragments. These trends arise due to
the simple but robust physics explained in \S \ref{sec:planets} -- to get
any planetary debris,  one needs to disrupt a gas fragment, and that process is more frequent at
low [M/H].

\subsection{Planet - debris disc relations}\label{sec:debris_planets}

\subsubsection{Gas giants}\label{sec:giants}

In our single migrating fragment simulations, just by the nature of the solid debris's origin, there cannot be a correlation between the presence of a debris disc and the presence of a gas giant. In fact, there is an anti-correlation: if a gas giant survives then it implies that no debris was released in a tidal disruption of a fragment. 

Before claiming that a clear anti-correlation is the prediction, let us reflect on the fact that in a multi-fragment protoplanetary disc the result is more complicated. For example, fragments born early on may migrate more rapidly and be disrupted, producing solid debris. Fragments appearing closer to the end of the disc's life may be migrating in slower and are more likely to survive. Therefore, one may have detectable giant planets and solid debris at the same time in this picture. However, there appears to be no reason to expect that a positive correlation between gas giants and planetary debris will arise in the multi-fragment system. The most likely outcome is that the debris disc presence will remain anti-correlated with the presence of gas giants, but the strength of this relation may be significantly diluted by the "noise" resulting from disruption of "other" fragments in the system that left behind some planetary debris.

It is encouraging that the qualitative anti-correlation trend between gas giants and the debris discs predicted by Tidal Downsizing appears consistent with contemporary observations. \cite{MarshallEtal14} finds that stars with observed gas giant planets are twice less likely to host debris discs as opposed to stars with lower mass ($M_{\rm p} \le 30 \mearth$ planets). 

\subsubsection{Sub-giant planets}\label{sec:sub-giants}

The single migrating fragment scenario also predicts a one-to-one correlation between the presence of debris and a lower-mass planet that was also formed due to the same gas fragment disruption.  However, a finite detectability of the debris discs and the planets will weaken this correlation, similarly to the arguments made in \S \ref{sec:giants}. In addition, tidal disruptions occur in the model at a broad range of separations, from sub-AU to tens of AU, and these spatial scales are accessible to debris disc observations, in general \citep{Wyatt08}. Observations of sub-giant exo-planets however probe only separations $a\lesssim $ a few AU with a few exceptions currently \citep[e.g.][]{HanEtal14}. Thus, for the planet to be observed it should also be able to migrate into that region before the protoplanetary disc is dissipated away.

Thus our model does predict a correlation between sub-giant planets and DDs, although it is weaker than the one-to-one relation that is appropriate {\em if all the debris discs and all the sub-giant planets}, irrespective of their mass and separation, were detectable. We may also expect a further reduction in the strength of the correlation in multi-fragment systems and also due to post-evolution of planetary and DD populations, which may see some of these be thrown out from the stellar system or destroyed completely. Future more sophisticated modelling is needed to address these issues. 


\begin{figure} 
\includegraphics[width=1.0\columnwidth]{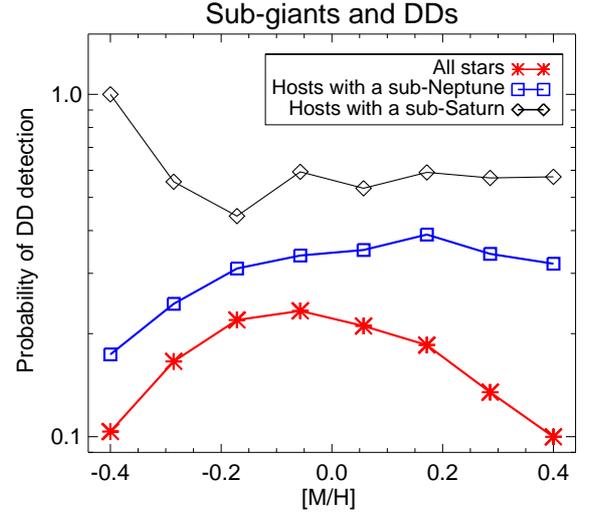}
\caption{Probability of a star with a given metallicity [M/H] hosting a detectable debris discs for all stars (red asterisks) and for stars that host either a sub-Neptune mass planet ($3 \mearth < M_{\rm p} < 15 \mearth$) or sub-Saturn one ($15\mearth < M_{\rm p} < 100 \mearth$), shown with blue squares and black diamonds, respecitvely. Presence of a sub-giant planet in the inner disc increases the chances that a detectable DD will be also present. }
\label{fig:bar_se_z_frac}
\end{figure}

\subsection{Preliminary comparison to observations}\label{sec:comparison}

\begin{figure} 
\includegraphics[width=1.0\columnwidth]{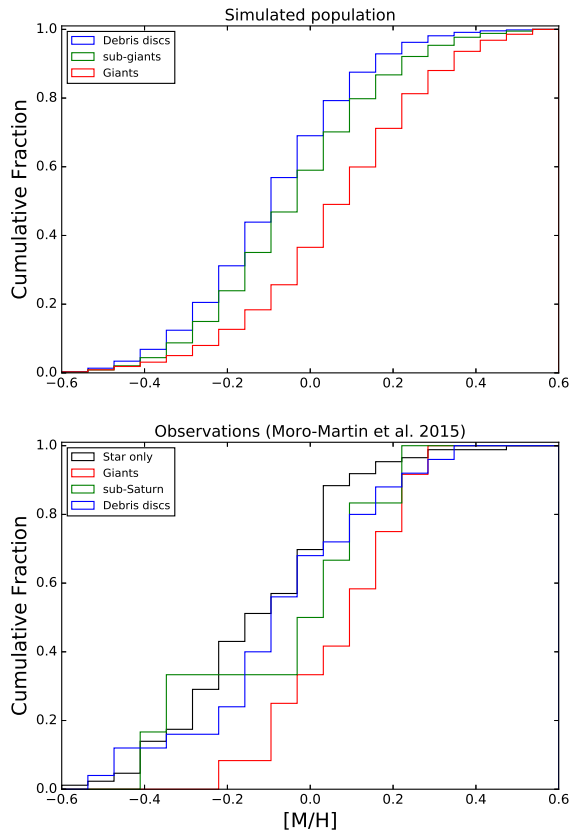}
\caption{{\bf Top panel:} the cumulative frequency distributions for debris discs, sub-giant and giant planets
over  metallicity of the host star obtained from our population synthesis calculations.  See the legend for the meaning of the histograms. {\bf Bottom panel:} data from a survey of nearby stars searching for DDs and planets \citep{Moro-MartinEtal15}.  Gas giant planets are clearly offset towards higher metallicity hosts in both the data and the simulations. Debris discs and sub-giant planets do not correlate with metallicity of the host.}
\label{fig:z_cdf}
\end{figure}

Fig. \ref{fig:z_cdf} shows how the cumulative fraction of systems with a DD or a planet is distributed over metallicity of the host stars, [M/H]. The top panel shows the results of this paper, whereas the bottom panel shows the data from \cite{Moro-MartinEtal15}.

The red histograms show the distribution of gas giants over [M/H]. In both the simulations and the observations these distributions are shifted towards higher metallicity. In TD, gas giants prefer high metallicity environments because pebbles (large grains) are more abundant at high [M/H], fuelling a higher pebble accretion rates onto gas fragments. Higher pebble accretion rates allow the fragments to collapse more rapidly \citep{Nayakshin15a}, so that a greater fraction of fragments born in the outer cold self-gravitating disc can reach the inner disc avoiding tidal disruption \citep{Nayakshin15b}.

The blue histogram in the top panel of fig. \ref{fig:z_cdf}  shows our theoretical debris disc distribution over [M/H], calculated for $M_{\rm det} = 0.5 \mearth$. This distribution is broad and has no obvious preference for high metallicity environs. The low panel of the figure shows that observations show a similarly wide [M/H] distribution for observed debris discs. Numerous previous observations confirm the insensitivity of DDs to the metallicity of the host star \citep[e.g.,][]{MaldonadoEtal12,MarshallEtal14}. Our results are hence qualitatively consistent with these observations.

The fact that the frequency of detection of super-Earths is also weakly dependent on [M/H] is well known \citep{MayorEtal11,HowardEtal12} and is evident in the bottom panel of fig. \ref{fig:z_cdf}. TD reproduces this result (see the green histogram in the top panel of the figure) with the physics already discussed above: at low [M/H], gas fragment disruptions are much more frequent than at high [M/H] but release less massive cores due to the relative scarcity of metals inside the fragments. This has been discussed previously in \cite{NayakshinFletcher15}.

\begin{figure} 
\includegraphics[width=1.0\columnwidth]{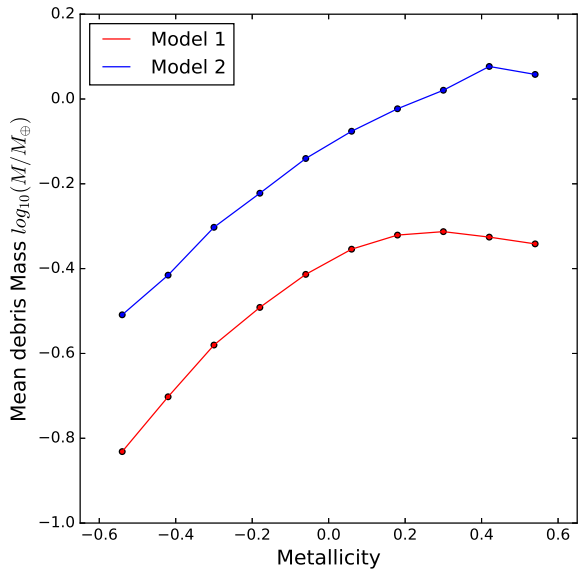}
\caption{Mean debris disc mass as a function of host star's metallicity for both model 1 and model
  2. These averages are taken from the raw data set with no detection limits
  or cuts applied. }
\label{fig:mean_mass_core}
\end{figure}

Very recently, \cite{GasparEtal16} have shown that the mean mass of a {\em detected} debris disc increases with [M/H] of the host star strongly. In fig. 
\ref{fig:mean_mass_core} we show the mean debris disc mass per disruption for model 1 and 2 for our population synthesis calculations. To reiterate, in model 1 the debris disc mass is proportional to the total metal content of the disrupted fragment, whereas in model 2 the debris mass is 10\% of the core mass. Clearly, despite the differences between these two assumptions, the DD mass versus metallicity trends of our calculations are qualitatively consistent with the observations reported in \cite{GasparEtal16}. The simpler model (1) has a flattening at [M/H]$\gtrsim0.3$, which may be caused by the feedback released by the massive cores onto their hosts in these high metallicity fragments. Indeed, model 2 curve demonstrates that the mean core mass continues to increase with [M/H], implying that the feedback effects are the strongest at high [M/H]. 

Fig. \ref{fig:mean_mass_core} is also qualitatively consistent with the fact that the brightest debris discs are only found around hosts with higher metallicities \citep[see][]{Moro-MartinEtal15,MontesinosEtal16}. We note again that multi-fragment population synthesis may quantitatively change the results but we expect the positive trend between debris disc mass and [M/H] to remain true.

%
%
%
%

\section{Discussion}\label{sec:discussion}

\subsection{Main results}\label{sec:list}

We found that trends in the debris disc mass with metallicity of the host star and with the likelihood of having a planet of a given type are very different in the Tidal Downsizing and Core Accretion  scenarios. This implies debris disc observations present us with a sensitive diagnostic of planet/debris formation theories.

The overarching result of our calculations is that the same physical effect -- pebble accretion -- can explain all of the observed metallicity correlations of planets {\em and} debris discs in the context of Tidal Downsizing. In particular,

\begin{enumerate}

\item Higher [M/H] hosts provide higher pebble accretion rates onto the fragments, causing them to collapse faster. This results in the larger fraction of gas fragments surviving tidal disruption and yielding a strong positive correlation between gas giants survived in the inner disc and the metallicity of the host.

\item The corollary to this is that few gas fragments are tidally disrupted at high [M/H], hence making debris ring formation events relatively rare compared to low [M/H] hosts. This explains why debris discs in TD cannot monotonically correlate with metallicity of the host star. The resulting debris disc (DD) correlation depends on the observational sensitivity of DD survey and can range from an anti- to a weak positive correlation (cf. fig. \ref{fig:bar_all_z_frac}).

\item The presence of a detected gas giant planet around a star implies that the fragment from which the planet originated did not go through a tidal disruption. In a single migrating fragment scenario, this would mean that gas giants and debris discs would be mutually exclusive. However, in a more realistic multi-fragment scenario, fragments other than the one that produced the observed giant planet could undergo tidal disruptions and produce debris. Therefore, we expect some anti-correlation between gas giants and DDs rather than a full incompatibility of the two populations. The exact relation between observed gas giants and debris discs as predicted by TD depends on the sensitivity of the survey.

\item Planets less massive than $\sim $ Saturn mass are also a result of tidal disruptions in the TD theory. Therefore, there should be a correlation between such planets and the DD presence (see fig. \ref{fig:bar_se_z_frac}). As explained in \S \ref{sec:sub-giants}, this correlation is one-to-one for a single fragment but may be diluted in a multi-fragment disc. It was found (fig. \ref{fig:bar_se_z_frac}) that sub-Neptune planets should correlate with debris disc presence weaker than more massive ones (those between Neptune and Saturn mass), although more work is needed to quantify this result.

\end{enumerate}

\subsection{Comparison to observations}

Our results, summarised in the list in \S \ref{sec:list}, are broadly consistent with the available observations:

\begin{enumerate}

\item Gas giant planets are well known to correlate with [M/H] of their host stars \citep{Gonzalez99,FischerValenti05}. 

\item Debris discs do not correlate with [M/H] of the host star, instead being abundant at all metallicities. Our results are consistent with this prediction for a DD detection threshold of $M_{\rm det}$ above $\sim 0.5\mearth$ (see fig. \ref{fig:bar_all_z_frac}). 

\item It came as a considerable surprise that DDs do not correlate with the presence of gas giant planets \citep{MMEtal07,BrydenEtal09,KospalEtal09}. In CA, these planets require the most metal-rich massive gas discs which are expected to have more massive debris discs left behind. In our scenario, however, DDs are expected to anti-correlate somewhat with gas giant's presence. \cite{Moro-MartinEtal15} finds that stars with detected gas giants are about twice less likely to host a debris disc than stars with smaller planets, showing that gas giants indeed tend to discourage DD formation.

\item It is currently not clear due to the limited statistics whether DDs correlate with the presence of less massive planets \citep{Moro-MartinEtal15}. However, some DD samples did show a weak correlation between DDs and low mass planets \citep{WyattEtal12}. This would be consistent with our calculations. More observations are needed to ascertain the correlations of DDs and planets better.

\item We also found that our model predicts an increase in the mean mass of the debris disc with metallicity of the host star, as found by \cite{GasparEtal16}. Related to that correlation, if only the brightest (most massive) debris discs are considered, then this result predicts a lack of bright debris discs at low [M/H] environments, as found by \cite{Moro-MartinEtal15,MontesinosEtal16}.

\end{enumerate}

\subsection{Predictions for future observations}\label{sec:future}

Although our calculations were performed for a fixed mass of the star, $M_* = 1 M_\odot$, we can speculate how these results may depend on the mass of the star. Assuming that initial fragment mass increases with the mass of the host star, we would expect relatively more disruptions of gas fragments at lower $M_*$ because low mass fragments are more susceptible to those disruptions than higher mass fragments \citep{NayakshinFletcher15}. We would therefore expect more debris per star around stars less massive than $1 M_\odot$ and less debris per star around more passive stars.

This is currently at odds with relative dearth of detected debris discs around low mass stars \citep[e.g.,][]{PlavchanEtal09}, but this may be an artefact of current surveys which are biased towards hotter earlier type stars \citep[see Introduction in][]{ChoquetEtal15}. Future observations should shed more light on the DD frequency trend with host star's mass.

%
%

\section{Conclusions}\label{sec:con}

Given the rarity of gas giant planets, the fact that presence of a giant at $\sim 1$~AU distance from the star somehow discourages formation of debris discs at distances of tens to hundreds of AU from the star is puzzling in the context of Core Accretion theory for planet formation (\S \ref{sec:dd_chal}). Non-correlation of DD presence with the host metallicity is also mysterious. The most well known success of CA theory, the prediction of the positive gas giant -- host star's metallicity correlation, assumes \citep{IdaLin04b,MordasiniEtal09b} that planetesimals are born much more readily at high [M/H].

In contrast, in Tidal Downsizing, the presence of a gas giant a few AU distance from the star signals that at least this fragment, born at $\sim 100$ or more AU from the star, was able to migrate inward all this distance without being tidally disrupted. This also means that the fragment did not release any large solids that it could have synthesised inside, thus making the presence of a debris disc less likely. A sub-giant planet (defined here as a planet less massive than Saturn) observed in the inner disc, however, is made in Tidal Downsizing by a disruption of a gas fragment further out, followed by migration closer in \citep{NayakshinFletcher15}. Presence of such a planet implies  that some debris has been actually produced. While multi-fragment discs will weaken these strong predictions (see \S \ref{sec:giants}), we still expect some antipathy between DDs and gas giants, as observed, whereas sub-giant planets should correlate to some degree with DD presence. Furthermore, as tidal disruptions are less likely at high [M/H], our model naturally explains why DDs do not correlate with host star's metallicity.

Future observations of DDs in different environs and their links to planets should help differentiate CA and TD scenarios of planet and debris formation.

%
%

\section{Acknowledgments}

Theoretical astrophysics research at the University of Leicester is supported
by a STFC grant. This research used the ALICE High Performance Computing
Facility at the University of Leicester. Some resources on ALICE form part of
the DiRAC Facility jointly funded by STFC and the Large Facilities Capital
Fund of BIS.

\bibliographystyle{mn2e}



\bsp	
\label{lastpage}
\end{document}